# High Pressure Effects on the Superconductivity in Rare Earth Doped CaFe$_2$As$_2$

Dan Cargill[1,2], Walter Uhoya[2], Krzysztof Gofryk[3], Georgiy M. Tsoi[2], Yogesh K. Vohra[2], Athena S. Sefat[3], and S. T. Weir[4]

[1]Department of Physics, Hendrix College Conway, AR 17302, USA

[2]Department of Physics, University of Alabama at Birmingham (UAB), Birmingham, AL 35294, USA

[3]Materials Science and Technology Division, Oak Ridge National Laboratory (ORNL), Oak Ridge, TN 37831, USA

[4] Mail Stop L-041, Lawrence Livermore National Laboratory (LLNL), Livermore, CA 94550, USA

**Abstract**: High-pressure superconductivity in a rare-earth doped Ca$_{0.86}$Pr$_{0.14}$Fe$_2$As$_2$ single crystalline sample has been studied up to 12 GPa and temperatures down to 11 K using designer diamond anvil cell under a quasi-hydrostatic pressure medium. The electrical resistance measurements were complemented by high pressure and low temperature x-ray diffraction studies at a synchrotron source. The electrical resistance measurements show an intriguing observation of superconductivity under pressure, with $T_c$ as high as ~51 K at 1.9 GPa, presenting the highest $T_c$ reported in the intermetallic class of 1-2-2 iron-based superconductors. The resistive transition observed suggests a possible existence of two superconducting phases at low pressures of 0.5 GPa: one phase starting at $T_{c1}$ ~48 K, and the other starting at $T_{c2}$~16 K. The two superconducting transitions show distinct variations with increasing pressure. High pressure low temperature structural studies indicate that the superconducting phase is a collapsed tetragonal ThCr$_2$Si$_2$-type (122) crystal structure. Our high pressure studies indicate that high $T_c$ state attributed to non-bulk superconductivity in rare-earth doped 1-2-2 iron-based superconductors is stable under compression over a broad pressure range.

**Introduction**

The interplay of high pressure and material chemistry has always played a pivotal role in the discovery and optimization of novel superconducting materials [1-3]. The discovery of superconductivity at 26 K in fluorine doped $LaFeAsO_{1-x}F_x$ by Hosono et al in 2008 [4-5] has generated a lot of interest in related layered ferro-pnictides [6-10]. In the resulting researches, bulk superconductivity has been obtained in several layered ferro-pnictides under application of high pressure or chemical doping upon the suppression of spin-density-wave (SDW) state. However, the maximum $T_c$ achieved to date in these compounds remains below 57 K achieved through electron doping in 1111 family [11] and 38 K achieved through hole doping in 122 family [12].

$CaFe_2As_2$ has the $ThCr_2Si_2$-type tetragonal (T) crystal structure at ambient conditions, and exhibits a tetragonal to orthorhombic structural transition and antiferromagnetic (AFM) ordering on cooling and a pressure-induced structural transition to a collapsed-tetragonal (CT) crystal structure at high pressures [1]. The structural and AFM transitions are suppressed under high pressures and superconductivity is induced with a $T_C$ maximum reported under high pressure [1]. Another way to induce isostructural tetragonal structure collapse is through chemical pressure [13, 14]. Saha et al. reported that even 7.5 % substitution of $Ca^{2+}$ with the rare-earth $Pr^{3+}$ is enough to cause the electron doped compound to collapse upon cooling before 51 K with higher doping concentrations leading to higher temperature of structure collapse [14].

Unusually non-bulk superconductivity characterized by a mysterious small volume fraction which superconducts up to $T_c$ onset = 42-49 K, much higher than the maximum $T_c \sim$ 38 K in 122 family, has been realized in single crystalline rare earth doped $Ca_{1-x}Pr_xFe_2As_2$ [10, 13]. Qi et al. reported superconductivity with $T_C$ up to 46 K for Pr doped $Ca_{87}Pr_{0.13}Fe_2As_2$ [15]. Lv et al. reported that all their $Ca_{1-x}Pr_xFe_2As_2$ samples doped between x=10.7% and 12.7% showed zero resistivity above 15 K with a superconducting onset as high as 49 K for x=12.7 % doped sample [13]. Saha et al. reported an onset $T_c$ reaching 47 K for 14% doping [14]. The $T_c$ of 45-49 K reported in Pr-doped $Ca_{1-x}Pr_xFe_2As_2$ at ambient pressure is the highest among the 122 family of Fe-based superconductors [10, 13-15]. Recently, it has been suggested that the nature of the filamentary, anisotropic superconductivity in $Ca_{0.86}Pr_{0.14}Fe_2As_2$ could be interfacial as evidenced from the resistive and magnetic results [10, 13]. However, no high pressures studies have been reported on the transport properties of the RE-doped compounds hence it's still unclear how application high pressure could further tune their structural, magnetic and superconducting properties. Therefore, it would be interesting to study how pressure affects the structural and interfacial superconducting properties of rare-earth doped $CaFe_2As_2$ compounds with optimal composition for high $T_c$. Pr doped $Ca_{0.85}Pr_{0.13\pm2}Fe_2As_2$ forms an ideal candidate for high pressure studies because it has the highest $T_c$ (49 K) reported to date among the 122 Fe-based materials [10, 13].

In this report, a series of temperature- and pressure-dependent electrical resistance measurements, up to ~ 12.2 GPa and down to ~11 K, were undertaken to investigate the pressure evolution of superconductivity in rare earth doped single crystalline $Ca_{0.86}Pr_{0.14}Fe_2As_2$, using a designer-diamond anvil cell (DAC). We further carried out simultaneous pressure- and temperature-dependent x-ray diffraction studies on the sample down to 9 K to determine the crystalline structure of the superconducting phase of $Ca_{0.86}Pr_{0.14}Fe_2As_2$. Our studies show that the single crystalline $Ca_{0.86}Pr_{0.14}Fe_2As_2$ exhibit a unique successive double non-bulk superconducting transition that is absent from other 122-iron based superconductors. The higher transition temperature, $T_{c1}$ is higher at lower pressures and decreases monotonically with pressure and could not be detected above 12 GPa for T >11 K. The lower transition temperature $T_{c2}$ is fairly constant and disappears between P= 3.2 and 4 GPa. Onset of upper transition ($T_{c1}$) reaches a record high superconducting transition temperature of 51±1 K which exceed the highest $T_c$ onset of bulk superconductivity ($T_c$~38) [12] recorded for 122 materials either by chemical doping or high pressures. Our room temperature x-ray diffraction studies show that the collapsed tetragonal phase induced in $Ca_{0.86}Pr_{0.14}Fe_2As_2$ by cooling to 62 K is stabilized at room temperature by pressures as low as 0.5±0.2 GPa. Our low temperature studies shows that the crystal structure of the high pressure superconducting phase is $ThCr_2Si_2$ type collapsed tetragonal (CT) structure and the sample remains in the CT phase even at extreme pressures (7.5±0.2 GPa) and low temperatures (9 K) where the sample still displays signatures of non-bulk high-$T_c$ superconductivity. This is the first report of high-pressure study of both structural and superconductivity in RE-doped $Ca_{1-x}Re_xFe_2As_2$.

**Experimental Details.**

Large platelets of single-crystal samples of $Ca_{0.86}Pr_{0.14}Fe_2As_2$ were grown from FeAs flux, similar to that described in reference [16]. The crystals were characterized using energy dispersive x-ray spectroscopy (EDS), x-ray diffraction (XRD), temperature-dependent magnetization and electrical resistance measurements. High-pressure and low-temperature electrical resistance measurements were preformed on the single crystal $Ca_{0.86}Pr_{0.14}Fe_2As_2$ were performed using four-probe method in a six-tungsten microprobe designer DAC with 280μm size cullet as described earlier [17-18]. The eight tungsten microprobes are encapsulated in a homoepitaxial diamond film and are exposed only near the tip of the diamond to make contact with the sample. The sample was loaded into a 165 μm hole of a 225 μm thick spring-steel gasket that was first pre-indented to a ~100 μm thickness and mounted between a matched pair of the 280 μm cullet size designer diamond anvils and 300 μm cullet size matching diamond, ready for high-pressure and low temperature experiments. Two electrical leads pass constant 5μA current through the sample and two additional leads measure voltage across the sample. Care was taken to electrically insulate the sample and the designer microprobes from the metallic gasket by using solid steatite as a pressure medium. In addition, the solid steatite pressure medium provides for a quasi-hydrostatic pressure measurement condition. Pressure was applied using a gas membrane to the designer DAC. For simultaneous temperature- and pressure-dependent x-ray diffraction experiments, the designer DAC was cooled down in a continuous helium flow-type-cryostat, and the pressure in the cell was measured *in situ* with the ruby fluorescence technique [18-19]. The synchrotron XRD experiments were performed at the high pressure beam-line 16-BM-D of HPCAT, at the Advanced Photon Source in Argonne National Laboratory. An angle dispersive technique with a MAR345 image-plate area detector was employed using a focused monochromatic beam with x-ray wavelength, $\lambda = 0.424600$ Å and sample to detector distance of 317.95 mm. The image plate XRD patterns were recorded with a focused x-ray beam of 5 μm by 7 μm (FWHM) on an 80 μm diameter sample mixed with ruby to serve as pressure marker. Experimental geometric constraints and the sample-to-image plate detector distance were calibrated using $CeO_2$ diffraction pattern and were held at the standard throughout the entirety of the experiment. The software package FIT2D [20] was used to integrate the collected MAR345 image plate diffraction patterns which were refined using GSAS [21] software package with EXPGUI interface [22] employing full-pattern Rietveld refinements and Le Bail fit techniques to extract structural parameters including atomic coordinates, bond angles, bond distances, lattice parameters, preferential orientation ratios.

**Results and Discussions**

Figure 1 shows the measured four probe electrical resistance of the $Ca_{0.86}Pr_{0.14}Fe_2As_2$ sample as a function of temperature at a fixed pressure of 0.5±0.2 GPa. The figure also illustrates the criteria used for defining the onset of upper transition temperatures of superconductivity ($T_{c1}$) and lower transition temperature of superconductivity ($T_{c2}$) for the sample. The sample pressure was monitored continuously by an in situ ruby fluorescence system during cooling and warming cycles to detect any changes. The pressure values reported are averages over the values recorded during the superconducting transition. The measured four probe electrical resistance shows two successive resistive transitions: an upper transition ($T_{c1}$) as shown by downturn in electrical resistance vs. temperature curve at the onset the upper transition. A second transition ($T_{c2}$) characterized by electrical resistance plateau followed by a sharper downturn occurs at lower temperatures relative to $T_{c1}$. The onset temperatures $T_{c1}$ and $T_{c2}$ is determined by the intersection of the two linear fits to the data above and below $T_{c1}$ and $T_{c2}$, as illustrated for P = 0.5 GPa in figure 1. The onset of resistive transition for ~0.5±0.2 GPa occurs at $T_{c1}$ = 48 K with lower transition occurring at $T_{c2}$ = 16 K.

Figure 2 shows temperature dependence of the normalized electrical resistance of $Ca_{0.86}Pr_{0.14}Fe_2As_2$ at various pressures. The upper transition ($T_{c1}$), and lower one ($T_{c2}$) are all noticeable for lower pressures (P < 3.9 GPa). Above 3.9 GPa, onset $T_{c1}$ gradually broadens and shifts to lower temperatures and eventually tend to vanish between 11.5 and 12.2 GPa for T > 11 K (figures 2, 3). The lower resistive transition $T_{c2}$ curve as a function of pressure behaves qualitatively different; $T_{c2}$ is fairly constant and is quickly suppressed below 11 K between 2.6 and 3.9 GPa, as compared to $T_{c1}$ which persists over a wider pressure range. Further increase in pressure beyond 11.5 GPa leads to a gradual increase in the overall sample resistance at low temperatures. The two step transitions at low pressures and the broad width of upper resistive transition above pressures where lower resistive transition could not be individually detected suggests a high possibility for the existence of two superconducting phases.

Since the present electrical resistance measurements are restricted to temperatures above 11 K, it is possible that the onset of $T_{c1}$ and $T_{c2}$ transition may occur below 11 K for pressures greater than 11.5 GPa and 4 GPa, respectively. Note that electrical resistance do not become completely zero in the superconducting phases for this sample in all pressure range investigated as is the case for other 122 materials under pressure studied previously by our group under similar quasi-hydrostatic pressure conditions in designer diamonds. The $T_{c1}$ of 48 K observed at ~0.5 GPa and the non-zero electrical resistance in the superconducting phase is consistent with recent reports on rare-earth $CaFe_2As_2$ at ambient pressure [10].

Figure 3 shows a clearer depiction of the evolution of the superconducting transition temperature with pressure. The onset temperature for the upper resistive transition $T_{c1}$ increases fairly in a narrower pressure range reaching a maximum of ~51±1 K at only 1.9 GPa and then decreases gradually with further increasing pressure up to ~11 GPa. The measured is typically 33-36 K lower than $T_{c1}$, and $T_{c2}$ decreases fairly linear as a function of pressure until a critical pressure of ~ 3.2 to 4 GPa above which it could not be detected. The measured $T_{c1}$ variation can be fitted by the following 3$^{rd}$ order polynomial equation over the entire pressure range, yielding $T_{c1}$ of 45.1 K at ambient pressure. The maximum $T_{c1}$ from the fit is at ~2.1 GPa and has a value of 51 K.

$$T_{c1} \text{ (in K)} = 45.11 \pm 1.03 + 5.79 \pm 0.73P - 1.65 \pm 0.14P^2 + 0.08 \pm 01P^3, \text{ (P is pressure in GPa)} \quad (1)$$

The $T_{c2}$ variation can be fitted by the following liner equation over the entire pressure range yielding 17.9 K at ambient pressure.

$$T_{C2} \text{ (in K)} = 17.87 \pm 0.73 - 1.23 \pm 0.39P, \text{ (P is pressure in GPa)} \quad (2)$$

Our measured value of $dT_{c1}/dP$ at ambient pressure (P≈0) is 5.79 K GPa$^{-1}$. This is higher in magnitude and different in sign as compared $dT_{c2}/dP$ of -1.23 K GPa$^{-1}$ reported for second superconducting phase confined to the lower pressure region between 0 and 3.2 GPa.

Our results show that the $T_{c1}$ dome for RE-doped CaFe$_2$As$_2$ is unusually broader extending over a wider pressure region as compared to typical pressure behavior of $T_c$ for Fe-based superconductors which show a drastic decrease with increasing pressure over a narrow pressure region [1]. Additionally, measured $T_{c1}$ values ranging from 45 to 51 K in the pressure range of 0 to 4 GPa are unusually higher for any known 122, 111, and 11 Fe-based superconductors under either chemical doping or pressure.

One possible explanation of the unique double superconducting resistive transitions, unusually high $T_c$ values and the broader $T_c$ vs. pressure dome is the existence of a low pressure minority phase in the Pr- doped Ca$_{0.86}$Pr$_{0.14}$Fe$_2$As$_2$, with tetragonal CaFe$_2$As$_2$ superconductor as one likely culprit. Given the layered nature of these materials; one can expect superconductivity to differently occur in the interface or intercalation places where doping is effective, forming numbers of superconducting sheets. These superconducting sheets consisting of low and high superconducting phases spread along the c-axis direction only take up a small percentage of the whole crystal. The structural properties of the phases can evolve differently with pressure resulting to different $T_c$ and magnetic properties evolution behavior. The low temperature resistive anomaly around 17 K at 0 GPa is suppressed by ~ 3.2 GPa and can be attributed to the small superconducting signal of the second phase. Previous studies show that the maximum $T_c$ for undoped CaFe$_2$As$_2$ under pressure is about 11-13 K [1] which is less than 17 K. Our electrical resistance measurement when the sample was decompressing to 2.6 GPa suggest that $T_{c1}$ is recovered at 47.5 K while the lower temperature phase could not be detected when pressure was released to 2.6 GPa. Additionally, Pr- doped $Ca_{0.86}Pr_{0.14}Fe_2As_2$ and $CaFe_2As_2$ are expected to have differing c-axis at ambient pressure and should be easily detected from diffraction studies. However, our x-ray diffraction study (see next sections) at ambient pressure and room temperature and at high pressures and low temperatures where the sample is superconducting does not show any evidence of secondary phases suggesting that a different possibility for the observed unusual trend from RE-doped $CaFe_2As_2$ exist.

Our optimal $T_c$ values of 51±1 K is consistent with that for the high temperature superconducting state for $Ca_{0.85}Pr_{0.13\pm2}Fe_2As_2$ with the maximum $T_{c1}$ of 49 K [10,13]. Chu et al has attributed the non-bulk Josephson junction array nature of the 49 K superconducting transition together with its unusually high enhanced $T_c$ to the probable long-sought-after interfacial mechanism for high temperature superconductivity proposed via exchange of excitons by Ginzburg in 1964 and Allender, Bray, and Bardeen in 1973 [23]. The origin for interfaces in a single crystalline sample has been determined by XRD to be misalignment of crystal grains which grows with doping along the FeAs-layers in Pr doped $Ca_{0.85}Pr_{0.13\pm2}Fe_2As_2$. The group proposed further that the microstructure induced by nano-crystallographic strains in Pr doped $Ca_{0.85}Pr_{0.13\pm2}Fe_2As_2$ samples provide the proper nano-morphology conducive to interfacial superconductivity responsible for enhancement of $T_c$ of the Pr-doped sample to 49 K. [10, 23].

Figure 4 shows representative x-ray diffraction patterns of $Ca_{0.86}Pr_{0.14}Fe_2As_2$ sample obtained at various temperatures and pressures. Figure 4(a) shows Rietveld refinement of the x-ray diffraction pattern at ambient temperature and 0.5 GPa using the $ThCr_2Si_2$-type (*I*4/*mmm*) crystal symmetry with the unit cell atom positions given as: Ca/Pr atoms at the 2*a* position (0, 0, 0), Fe atoms at the 4*d* positions (0, 1/2, 1/4) and (1/2, 0, 1/4), and As atoms at the 4*e* positions (0, 0, *z*) and (0, 0, -*z*), and refined value of z given in Table 1. The crystallographic parameters including lattice parameters, unit cell volume, bond angles and distances obtained from the refinement of the XRD pattern at 0.5 GPa are summarized in Table 1. The diffraction lines are all identified and indexed and the difference between the observed x-ray diffraction pattern and Rietveld fit is satisfactory small suggesting a single phase tetragonal crystalline sample with $ThCr_2Si_2$ similar to earlier reports on parent 122 materials [1, 3, 6]. The As-As bond distance of 3.1059 Å obtained at 300 K and 0.5 GPa is nearly equal to critical value of ~ 3 which was determined to mark the onset of tetragonal to collapsed tetragonal (T-CT) structural transition in related 1-2-2 materials [14].

Figure 4 (b) shows the x-ray diffraction pattern of the superconducting $Ca_{0.86}Pr_{0.14}Fe_2As_2$ at 9 K and 7.5 GPa where the sample is expected to be superconducting with a $T_{c2}$ onset of ~28 K as depicted in figure 3. The low temperature XRD pattern is similar to the room temperature pattern suggesting that the symmetry is also tetragonal. The Rietveld refinements of the x-ray diffraction pattern (figure. 1 (b) confirmed a $ThCr_2Si_2$ type tetragonal crystal structure with lattice parameters $a$ = 3.9971 Å, $c$ = 10.1491 Å, volume V=162.1463 Å$^3$, an axial ratio ($c/a$) =2.5391, atomic parameter z = 0.361356, As-Fe-As bond angle α=104.023(9) degrees and As-As bond distance of 2.8142(10) Å. Comparison the values obtained at 9 K, 7.5 GPa with the low pressure and room temperature values show an increase in *a*-lattice parameter while *c*-lattice parameter is strongly reduced suggesting a formation of the isostructural tetragonal collapsed phase. The As-As interlayer separation was determined to be the key parameter controlling the isostructural tetragonal structure collapse in 122 materials: $CaFe_2As_2$ crystal collapse once the interlayer As-As distance reaches a critical value of ~ 3 Å where As *p*-orbitals overlap upon application high pressure or rare-earth doping [14]. The measured As-As distance for our sample at 9 K and 7.5 GPa is 2.8142(10) Å confirming that the high-$T_c$ superconducting phase remains in the collapsed structure phase at higher pressures. Full structural refinement data for $Ca_{0.86}Pr_{0.14}Fe_2As_2$ at 9 K and 7.5 GPa in the stabilized collapsed phase are tabulated in Table I. It is believed that $T_C$ values are maximum when the corrugated FeAs layers exhibit coordination close to the ideal tetrahedral configuration (α = 109.9). Applied pressure drives the structure away from this ideal tetrahedral coordination by 3.4 % at 0.5 GPa where $T_c$ is 48 K and by 5.35% at 7.5 GPa in the CT phase.

**Conclusions**

In summary, pressure and temperature-dependent electrical resistance measurements are reported on $Ca_{0.86}Pr_{0.14}Fe_2As_2$ using a designer DAC to 12.2 GPa and down to 11 K. The resistance measurements show an evidence of a pressure-enhanced superconductivity with a remarkable maximum $T_{c1}$ of ~51 K at only 1.9 GPa. The onset of superconducting transition temperature decreases with increasing pressure and disappears for pressures above ~12 GPa above 11 K. The complementary x-ray diffraction studies conducted under high pressure and low temperature conditions suggest that that the crystal structure of the superconducting phase of the rare rare-earth doped $Ca_{0.86}Pr_{0.14}Fe_2As_2$ is of the collapsed tetragonal $ThCr_2Si_2$-type. Our results show that for certain levels of rare earth doping into 122 compounds, pressure may be a viable means of further optimizing $T_c$ and transition temperatures as high as 51 K or more may be obtainable. Our high pressure studies indicate that the high-$T_c$ superconductivity in RE-doped 1-2-2 iron-based superconductors is stable over a broad pressure range.

**Acknowledgment**

Dan Cargill acknowledges support from the National Aeronautics and Space Administration (NASA)-Alabama Space Grant Consortium, National Science Foundation (NSF) Research Experiences for Undergraduates (REU)-site under grant no. NSF-DMR-1058974 awarded to UAB. Walter Uhoya acknowledges support from the Carnegie/Department of Energy (DOE) Alliance Center
(CDAC) under grant no. DE-NA0002006. The work at ORNL was supported by the Department of Energy, Basic Energy Sciences, Materials Sciences and Engineering Division. Portions of this work were performed in a synchrotron facility at HPCAT (Sector 16), Advanced Photon Source (APS), Argonne National Laboratory. We are grateful to J. E. Mitchell for growing the single crystal of $Ca_{0.86}Pr_{0.14}Fe_2As_2$.

**Figure Captions**

**Figure 1**. (Color online). Temperature dependence of the electrical resistance of rare earth doped $Ca_{0.86}Pr_{0.14}Fe_2As_2$ at 0.5 GPa. Steatite was used as pressure medium and pressure was calibrated using ruby fluorescence.

**Figure 2**. (Color online). Temperature dependence of the electrical resistance of rare earth doped $Ca_{0.86}Pr_{0.14}Fe_2As_2$ at various applied pressures. Steatite was used as pressure medium, and pressure was calibrated using ruby fluorescence.

**Figure 3**. (Color online). Measured superconducting transition temperature for $Ca_{0.86}Pr_{0.14}Fe_2As_2$ as a function of pressure, by onset temperature criterion shown as $T_{c1}$ and $T_{c2}$. The solid curves are third order polynomial fit to $T_{c1}$ data and liner fit to $T_{c2}$ data, (see text for detail).

**Figure 4.** (Color online) Rietveld refinement of powder x-ray diffraction patterns of $Ca_{0.86}Pr_{0.14}Fe_2As_2$ in the $ThCr_2Si_2$ (I4/mmm) crystal phase at (a) ambient temperature and 0.5±0.2 GPa, and at (b) 9 K and 7.5±0.3 GPa. The lowermost solid line (magenta) in (a) and (b) are the difference profile curves between the observed (solid red symbols) and calculated (green line) profiles. The hkl values for peaks corresponding to the tetragonal *I*4/*mmm* phase are marked by the black lines. The crystallographic parameters obtained from GSAS refinement are summarized in Table 1.

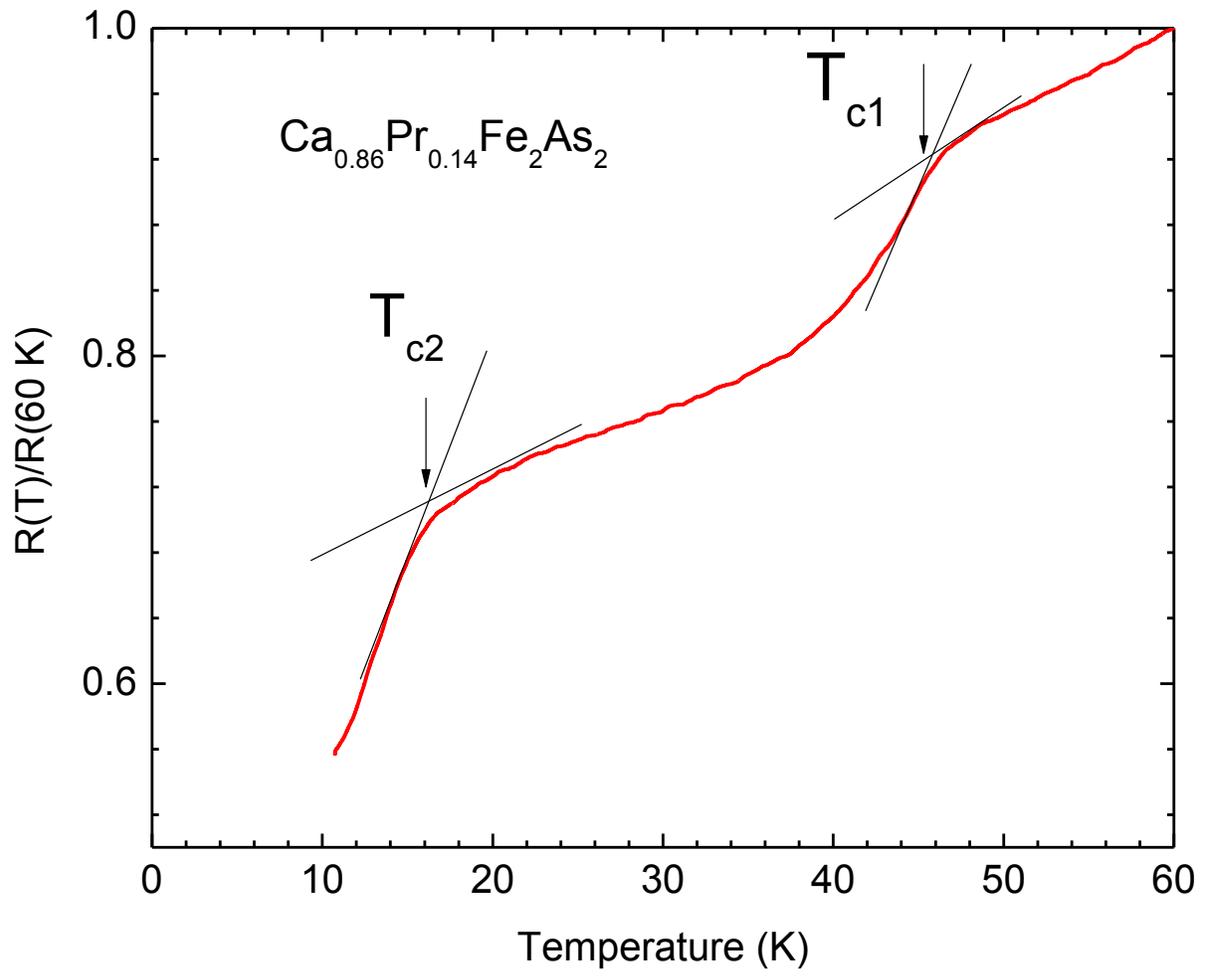

Figure 1

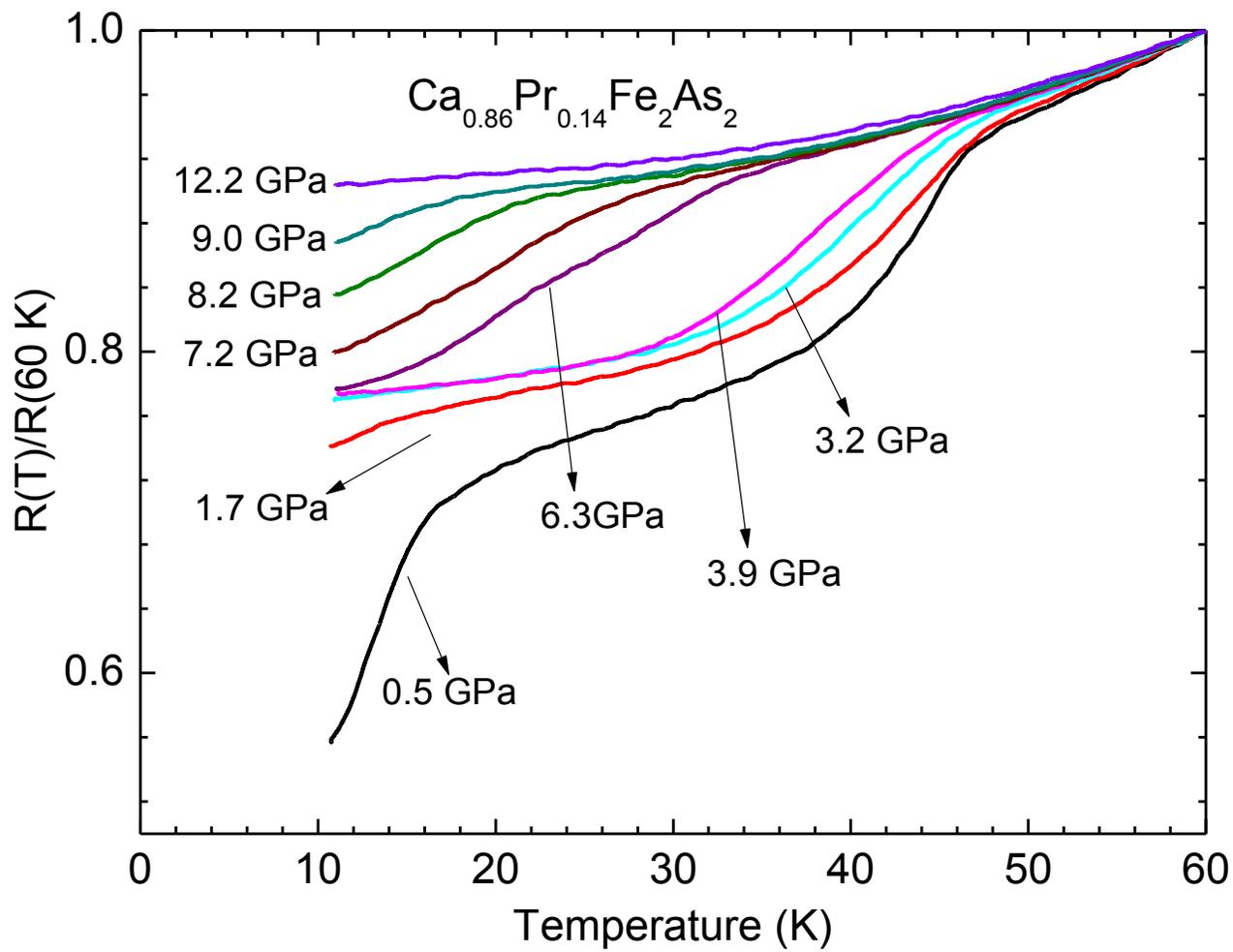

Figure 2

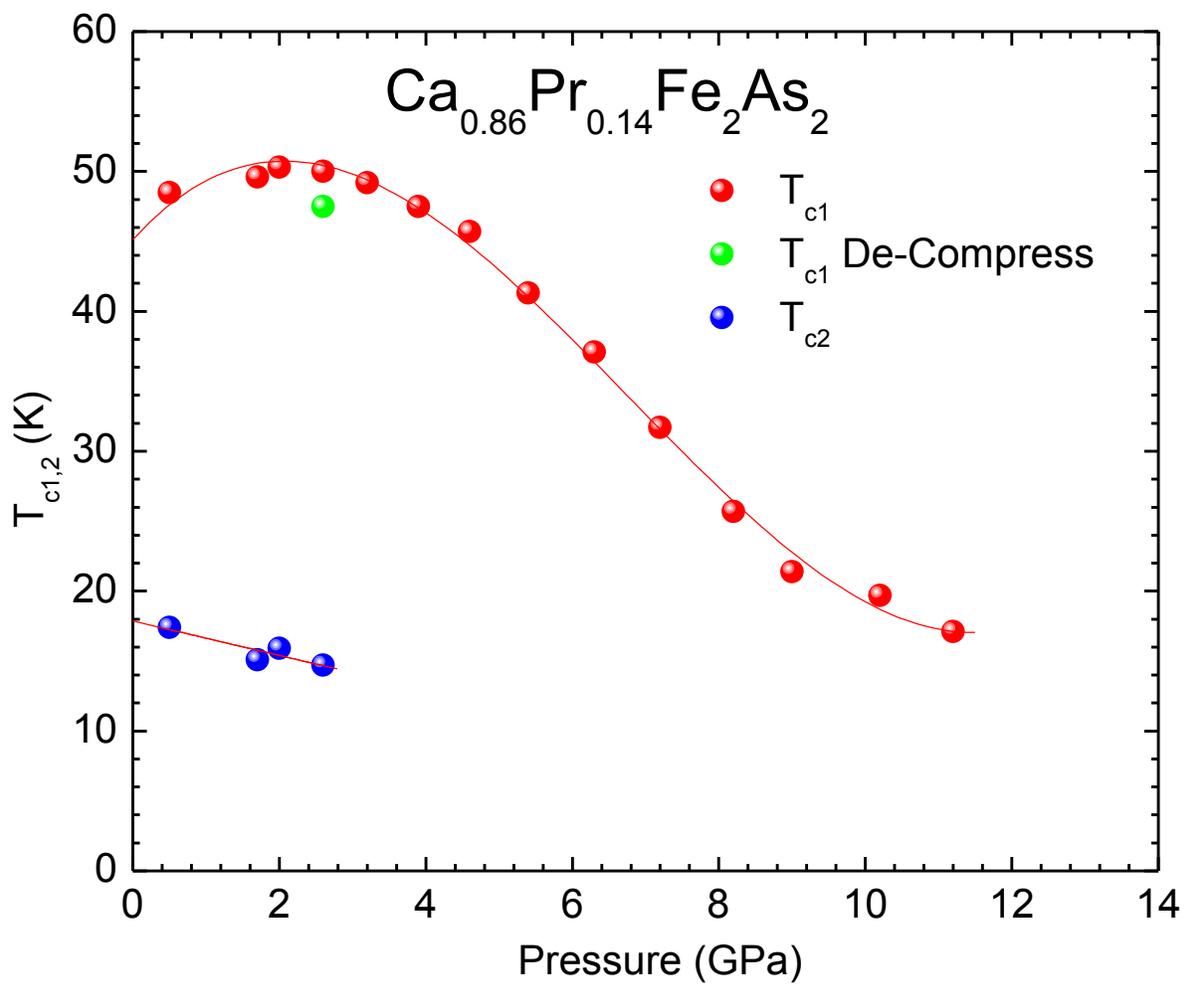

Figure 3

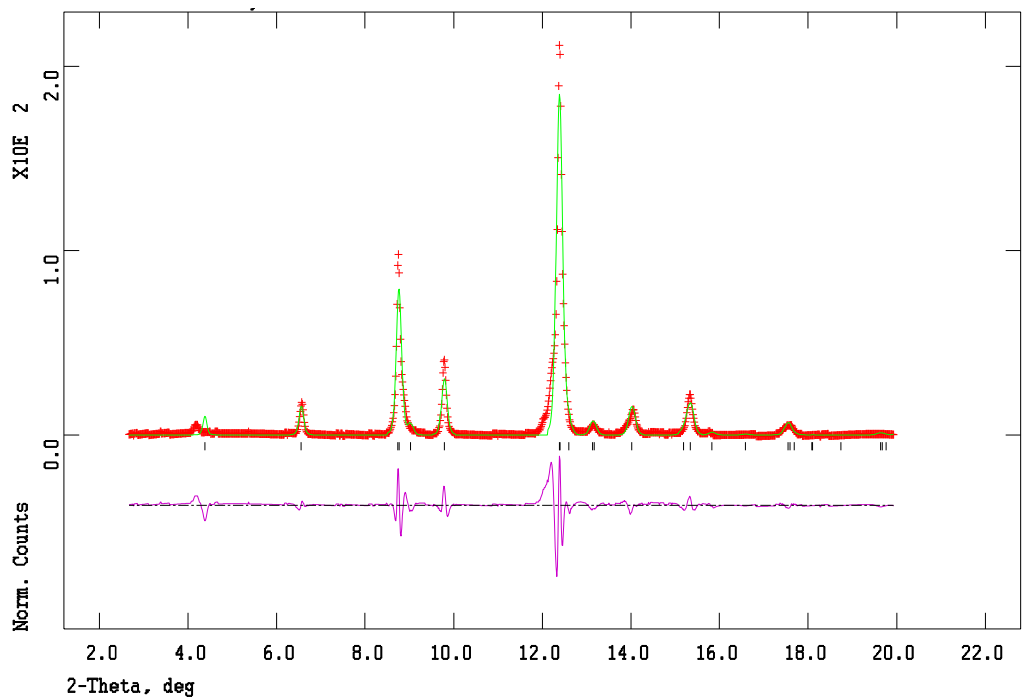

Figure 4 (a)

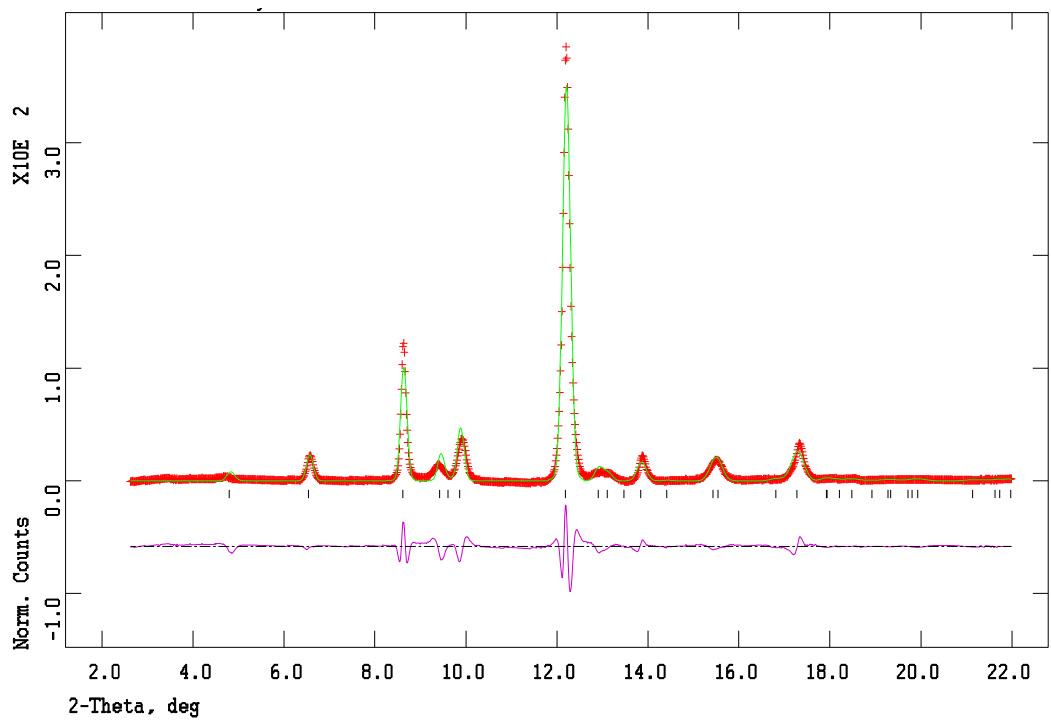

Figure 4 (b)

**TABLE I**. Crystallographic data for $Ca_{0.86}Pr_{0.14}Fe_2As_2$ crystals as determined from GSAS refinement of synchrotron x-ray diffraction for selected pressures and temperatures at which the sample adopt the collapsed tetragonal $ThCr_2Si_2$ (122) crystal structure. Atomic positions for space group I4/*mmm* are: Ca/Pr at 2*a* position (0, 0, 0); Fe at 4*d* positions (0, 1/2, 1/4) and (1/2, 0, 1/4); and As at 4*e* positions (0, 0, z) and (0, 0, -z). The measurements were carried out using a focused x-ray monochromatic beam of wavelength 0.4246 Å. Sample shows preferred orientation along (001) with a GSAS refinement March-Dollase Preferential Orientation ratio of 2.45539 at (9 K, 7.5 GPa) and 2.382975 at (0.5 GPa, 300 K).

| Temperature | 300 K | 9 K |
|---|---|---|
| Pressure | 0.5 GPa | 7.5 GPa |
| Space Group(phase) | I4/mmm | I4/mmm |
| Lattice parameter, $a=b$ (Å) | 3.936922(1) | 3.99705(63) |
| Lattice parameter, $c$ (Å) | 11.106790(2) | 10.14911(652) |
| Unit cell volume, $V$ (Å$^3$) | 172.1481 | 162.1463 |
| Atomic parameter, (z,-z) | 0.360178, 0.639822 | 0.361356, 0.638644 |
| Ca/Pr U-iso | 0.04156 | 0.04146 |
| Fe U-iso | 0.11294 | 0.07555 |
| As U-iso | 0.07759 | 0.08304 |
| GSAS wRp | 0.1049 | 0.1250 |
| GSAS Rp | 0.0650 | 0.0807 |
| Reduced Chi ^2 | 0.2630 | 0.5993 |
| Bond Angles (deg) | | |
| As-Fe-AS ($\alpha$) | 106.1859(11) | 104.023(14) |
| As-Fe-AS ($\beta$) | 116.2644(15) | 121.024(32) |
| Fe-As-Fe | 73.8148(8) | 75.977(14) |
| Bond lengths (Å) | | |
| As-As | 3.1059(10) | 2.8142(18) |
| Fe-As | 2.3178 (9) | 2.2959(4) |
| Fe-Fe | 2.78382(33) | 2.8263(4) |